\documentclass[12pt,preprint]{aastex}
\usepackage{emulateapj5}
\def\gtrsim{\mathrel{\hbox{\rlap{\hbox{\lower4pt\hbox{$\sim$}}}\hbox{$>$}}}}
\let\ga=\gtrsim
\def\lesssim{\mathrel{\hbox{\rlap{\hbox{\lower4pt\hbox{$\sim$}}}\hbox{$<$}}}}
\let\la=\lesssim
                                
\begin{document}

                                \title{
Direct Measurements of Gas Bulk Flows in the ICM of the Centaurus Cluster with the {\sl Chandra} Satellite
                                }

                                \author{
Renato A. Dupke \& Joel N. Bregman
                                }

                                 \affil{
University of Michigan, Ann Arbor, MI 48109-1090
                                }

%===============================================================================
                                \begin{abstract}
We present the analysis of the velocity structure of 
the intracluster gas near the core of Abell 3526 obtained with two off-center Chandra
observations, specifically designed to eliminate errors due to spatial variations of the 
instrumental gain. We detected a significant velocity gradient along the direction NE-SW direction, 
roughly perpendicular to the direction of the incoming sub-group Cen45, in agreement 
with previous ASCA SIS measurements. 
The presence of gas bulk velocities is observed both  
with and without the inclusion of the FeK line complex in the spectral fittings.
The configuration and magnitude of the velocity gradient is consistent with near 
transonic circulatory motion, either bulk or eddy-like. 
The velocity difference obtained using the best calibrated central regions of ACIS-S3 
is found to be (2.4$\pm$1.0) $\times$10$^{3}$ 
km s$^{-1}$ for rectangular regions 2$^\prime$.4$ \times $3$^\prime$ roughly diametrically opposed 
around the cluster's core. There are also indications of a high velocity zone towards the 
Southern region with similar magnitudes. The detection of velocity gradients is significant at the
$>$99.4\% confidence and simulations show that intrachip gain fluctuations $>$ 1800 km s$^{-1}$ are
required to explain the velocity gradient by chance.
The measurements suggest that $>$1\% of the total merger energy can still be 
bulk kinetic 0.4 Gyr after the merging event.  
This is the first direct confirmation of velocity gradients in the intracluster 
gas with independent instruments and indicates that strong departure from hydrostatic equilibrium 
is possible even for cool clusters that do not show obvious signs of merging.

                                \end{abstract}

%===============================================================================
                                \section{
Introduction
                                }

Clusters of galaxies form by the infall/merging of smaller scale 
systems. The merging of sub-clusters or groups of galaxies produces 
many  intracluster gas signatures, such as temperature and density inhomogeneities and
 gas bulk motions, all of which are observables in X-rays. 
The comparison of temperature and surface brightness 
distributions to numerical simulations of mergers can tell us about the 
evolutionary stage of clusters 
(e.g. Evrard 1990; Katz \& White 1993; Roettiger, Burns \& Loken 1993,1996; 
Pearce, Thomas \& Couchman 1994; Navarro, Frenk \& White 1995; Evrard, Metzler, \& Navarro 1996). 
In particular, simulations predict that long lasting gas bulk motions and turbulence 
will appear often as a consequence of cluster merging
(e.g. Ricker 1998; Roettiger, Stone, \& Mushotzky 1998; Takizawa \& Mineshige 1998; Burns et al. 1999; 
Takizawa 1999, 2000; Roettiger \& Flores 2000; Ricker \& Sarazin 2001; Pawl, Evrard \& Dupke 2005). 
Since the intracluster medium (ICM) is enriched with heavy elements,
gas velocities can be detected through Doppler shift of spectral lines 
(Dupke \& Bregman (2001a,b), hereafter DB01a,b), line broadening (Sunyaev, Norman \& Bryan 2003; Pawl et al. 2005)
and also through kinetic S-Z effect using Bolometers (Dupke \& Bregman 2002; Sunyaev et al. 2003).

Previous X-ray studies of the physical state of the ICM have concentrated 
mostly on the analysis of temperatures,
density and metal abundance distributions. In order to zero in the precise evolutionary 
history of galaxy clusters and to
determine the level of departure from hydrostatic equilibrium, 
it is very important to take into account gas velocities, a critical diagnostic that has largely 
been missing from cluster analysis, and that provides the most direct information about the dynamics of 
the intracluster gas. 
The X-ray analysis of intracluster gas bulk velocities became technically feasible only 
recently, after the calibration
of the instrumental gain of the {\sl ASCA} spectrometers was completed. 
The first direct detections of intracluster gas bulk velocities were found for the Perseus and Centaurus 
clusters (DB01a,b) using {\sl ASCA} data. 
In the case of Centaurus, a relatively high velocity gradient ($\sim$3000 km~s$^{-1}$) was detected 
near the clusters's core 
with the SISs, which are the {\sl ASCA} spectrometers with best gain stability and spectral resolution.

Given the importance of velocity gradients to understand the dynamics of the intracluster gas
it is crucial to confirm and 
improve the measurements of bulk velocities in these clusters. The {\sl Chandra} satellite provides an excellent 
opportunity for this task. It has a good gain temporal stability (Grant, C. 2001) and
we can eliminate the spatial variations of the intrachip gain by performing spatially resolved spectroscopy 
of multiple cluster regions using the {\it the 
same CCD location}. In this Letter, we present results of a double pointing observation 
of Centaurus specifically designed for bulk gas velocity measurements. 

%===============================================================================

\section{Previous Velocity Studies on Centaurus}

Centaurus (Abell 3526) is a Bautz-Morgan type I, it has an optical redshift 
of 0.0104 and is one of the closest X-ray bright clusters of galaxies. The temperature of the 
gas in the cluster is $\sim$ 3.6 keV (e.g. Peres et al. 1998) and decreases towards the 
center to $\sim$ 1 keV (Sanders \& Fabian 2002). {\sl ASCA} GIS \& SIS 
analyses of the central region of Centaurus have 
shown evidence of a strong central metal abundance enhancement (Fukazawa et al. 1994)
varying from supersolar near the 
central regions down to 0.3 solar at $\sim$ 13$^\prime$ (1$^\prime$ at Centaurus distance is equivalent
to $\sim$ 19 h$^{-1}_{50}$kpc). 
At very small radii ($\la$0$^\prime$.5--1$^\prime$) there is significant substructure seen in surface brightness 
(plumes),
 temperature, and metal abundances (Sanders \& Fabian 2002; Fabian et al. 2005). 
At large radii the cluster shows signs of interaction (Churazov et al. 1999) 
with a sub-group (Cen 45) discovered optically (Lucey, Currie \& Dickens 1986a,b; Stein et al. 1997). 
The main group (Cen 30) which is centered on the cD galaxy (NGC 4696) shows an average radial velocity of 
$\sim$3400 km~s$^{-1}$ and a velocity dispersion of $\sim$900 km~s$^{-1}$. It is accreting the Cen 45 group, 
which is associated with the galaxy NGC 4709 at $\ga$15$^\prime$ from NGC 4696 and has an average radial velocity
of $\sim$ 4700 km~s$^{-1}$ and a velocity dispersion of $\sim$ 150 km~s$^{-1}$. The velocity difference between 
Cen 45 and the main body of Centaurus found in the optical frequencies ($\sim$1500 km~s$^{-1}$) was also corroborated 
in X-rays (DB01b).  DB01b also used the SIS on {\sl ASCA} to study the intracluster 
gas properties at intermediate spatial scales ($\sim$3$^\prime$--8$^\prime$). They found a strong velocity gradient 
 in both SIS0 \& SIS1. The maximum velocity difference was found to be (3.3$\pm$1.1)$\times$10$^3$ km~s$^{-1}$ along the 
direction perpendicular to that of the incoming subgroup Cen 45 and they argued that it is likely due to a 
prior merging event with a strong line of sight component in the Centaurus cluster.

%===============================================================================
\section{Data Reduction and Analysis}

Abell 3526 off-center pointings analyzed in this work were observed using {\sl Chandra} 
ACIS-S3 in April 18th 2003, for 35 ksec each. After removal of high background periods we were left with 
34.3 ksec \& 33.9 ksec for the two pointings. 
Both observations were performed consecutively. 
In both cases the CCD was centered on the regions of expected maximal departure from the 
systemic velocity as derived from a previous {\sl ASCA} analysis (DB01b). 
The spatial configuration of the regions analyzed in this work and the relative location 
of the cluster's centers with respect to the ACIS-S3 chip are shown in Figures 1.
Here, we focus on the results of the central CCD regions, that are at the same CCD coordinates 
but still far enough from the spectroscopically complex cluster core. The use of the same 
CCD coordinates to study the regions of interest allows us to 
improve the accuracy of the 
measured gas velocities since the intrachip gain variations are minimized.

We used Ciao 3.2.2 with CALDB 3.0.3 to screen the data.
The data were cleaned using the standard procedure\footnote{http://cxc.harvard.edu/ciao/guides/acis\_data.html}.
Grades 0,2,3,4,6 were used. ACIS particle background was cleaned as prescribed for VFAINT mode. 
A gain map correction was applied together with PHA and pixel randomization. Point 
sources were extracted and the background used
in spectral fits was generated from blank-sky observations using
the {\tt acis\_bkgrnd\_lookup} script. 
Here we show the results of spectral fittings with XSPEC V11.3.1 (Arnaud 1996) using an absorbed 
{\tt VAPEC} thermal emission models. 
Metal abundances are measured relative to the solar photospheric values of 
Anders \& Grevesse (1989).
Galactic photoelectric absorption (NH) was incorporated using the {\tt wabs} 
model (Morrison \& McCammon  1983).
Spectral channels were grouped to have at least 20 counts/channel. Energy 
ranges were restricted to 0.5--8.5 keV. The spectral fitting parameter errors 
are 1-$\sigma$ unless stated otherwise.

There is a known 
reduction of quantum efficiency due to molecular contamination build up on optical blocking filter
(or on the CCD chips)\footnote{http://cxc.harvard.edu/cal/Links/Acis/acis/Cal\_prods/qeDeg/index.html}, which 
affects the count rates significantly at energy ranges below 1 keV. This also affects determination of 
low energy lines abundances, hydrogen column density and overall gas temperature. Previously corrected with
the {\it ACISABS} model or {\it corrarf} routine, this correction is now done automatically within the
{\it mkwarf} routine. The use of the newly incorporated routines reduced significantly the 
intrachip scatter of recovered value of NH across the detectors. The overall value of NH was 
near (within 10\%) Galactic and we, therefore, fixed it at the Galactic value (8$\times$10$^{20}$cm$^{-2}$) 
for all spectral fittings presented here.
We will refer to the observation where {\sl ASCA} found low velocities 
(NE pointing) as ``MINUS'' (Figure 1-Left) and the observation where {\sl ASCA} found high velocities 
 (SW pointing) as ``PLUS'' (Figure 1-Right). Also, in Figures 1 we
indicate the regions used in this work for spectral extraction of each CCD on the X-image. We
also overlay the X-ray contours showing the central regions and the CCD border.

 \subsection{
 ACIS-S3 Temporal and Spatial Gain Stability
                                 }

In order to estimate the level of global temporal gain fluctuations in both observations for the 
regions of interest, we divided each event file in 7 different time intervals, with near 5 ksec
in each interval. We used a region corresponding spatially to the full CCD 
with $\sim$ 1$^\prime$ borders excluded. We then 
extracted spectra and determined the best-fit parameters.  We show the results in Figures 2a
for the redshifts. It
can be seen that the
``PLUS'' pointing has been affected by stronger fluctuations than the ``MINUS'' pointing. 
The ``MINUS'' pointing shows a significantly smaller overall redshift scatter. 
The temporal variation of gain increases the overall uncertainties of the velocity analysis 
differently in each pointing. We included in quadrature the standard deviation 
of the gain scatter over time in the final estimation of the velocity gradients.

 We can also estimate the change in intrachip gain fluctuations within the observation 
 by looking at the velocity change in different positions of the CCD. Using the same event 
 files obtained for the 7 different epochs as those described in the previous paragraph, we determined 
 a velocity map through an adaptive smoothing routine that keeps a fixed minimum number of counts
 per region (here we used 5000 counts) to maintain the range of fitting errors more or less 
 constant for different regions. We then 
 determined the standard deviation of the best fit velocities for the same region through 
 different time periods. We plot the results in Figures 2b,c, where regions of high scatter are brighter.
  The color steps in Figures 2b,c represent the average 1$\sigma$ fitting
 errors of the individual regions used to construct the velocity map. 
 Given the limited photon statistics the precision with which we can determine the 
 variations is relatively poor. Nevertheless, it can be seen that the ``PLUS'' observation is also more 
 affected by intrachip temporal gain variations, especially near the SE, partially including the {\bf R+1} region. 
 In the `MINUS'' side the region most affected by the intrachip temporal gain fluctuation is {\bf R-1}. Overall, the 
 regions chosen for velocity analysis , i.e. along the strip (see below), {\bf PEAK-} and {\bf PEAK+} do not
 encompass regions of very high gain instability.

In order to check for any global gain shift between observations, we used the part 
of the CCDs that corresponds to the same ``sky'' regions. We call 
them overlap regions and denote them by OV- (in the ``MINUS'' pointing) and OV+ (in the ``PLUS'' 
pointing) in Figures 1. We present the results of the redshift variation for different cases
(different spectral ranges, with core included/excluded) in Table 1.
It can be seen in Table 1 that there are significant changes in the best fit values for velocities in the
overlapping regions of both pointings. The discrepancy between different pointings can be as high as 
$\sim$ 3 $\times$10$^3$ km~s$^{-1}$. However this extreme value only happens when the 
cluster core region is included and the soft energy band ($<$ 1 keV) is used in the spectral 
fittings. Given that the $\chi^2$ values of these fittings are unacceptably high 
it is clear that a simple VAPEC cannot 
accurately describe the physics of the cluster core and its inclusion is biasing the redshift measurements
(see DB05 for a description of redshift biases due to spectral modelling). 
Furthermore, direct fitting the iron lines with Gaussians produces results for the velocity differences 
similar to those obtained through spectral fittings when the core is excluded. For the FeL complex 
(from 0.95--1.25 keV) a best fit difference of (1.12$\pm$0.8)$\times$10$^3$ km~s$^{-1}$ is found, while for
the FeK complex a value of  (1.47$\pm$0.54) $\times$10$^3$ km~s$^{-1}$.  

The error weighted average 
(spectral fits without the core, FeL and FeK
complex) for the gain correction in between observations is (1.23$\pm$0.09)$\times$10$^3$ km~s$^{-1}$.
This global gain shift is most likely due to residual 
errors in the default correction for a non-uniform QE degradation that are still present in the new calibration, 
and also because the overlapping regions do not overlap in ``CCD'' coordinates and are  
at opposite locations with respect to the frame store. If we applied this global gain shift to the 
whole CCD the result would be a ``boost'' of the velocity gradient by $\sim$ 1.2$\times$10$^3$ km~s$^{-1}$. 
However, given the uncertainties for intrachip gain fluctuations we
conservatively do not use this correction to boost the redshifts of the 
``PLUS'' region but we point out that the velocity gradient that we obtain can be seen as a lower limit. 

 \section{
Velocity Distribution Along the CCD Targeted Regions
                                 }
The main advantage of the observational strategy used here is that we can use the same regions in CCD coordinates 
to analyze the ``sky'' regions previously suspected of having discrepant velocities. By using the 
same CCD region we can, in principle, eliminate spatial gain variations. Temporal variations of  
gain are also minimized by performing the two observations consecutively. The residual gain 
variations between the two observations can also be estimated using the 
overlapping regions as shown in the previous section. The final velocity measurements should only be affected by 
short term random local (subregions inside the CCD) gain variations. 

We chose a strip of similar rows near the CCD center to perform a detailed 
spectroscopic analysis of gas velocities. This choice was based on the degradation of quantum efficiency with 
chip row\footnote{cxc.harvard.edu/cal/Acis/Cal\_prods/qeu/qeu.pdf} and also on 
 the intrinsic spectral differences of the ICM between regions near the cluster's core and regions $>$2$^{\prime}$ 
 from the center. The ``strip'' is the most reliable region because it avoids the clusters' core 
 and also regions at very different distances from the frame store. It also covers the most
 stable region for intrachip gain (Figures 2b,c). Due to the 
surface brightness asymmetric distribution and the higher uncertainties involved in the ``PLUS'' dimmer pointing we 
chose the size of the regions to encompass at least 15 kcounts each. In the ``MINUS'' observation we show the results 
for spectral fittings using  similar sizes and in general they have twice as many counts as their counterparts 
on the ``PLUS'' side (22, 30 and 30 kcounts for {\bf R-1}, {\bf R-2} and {\bf R-3}, respectively). The same 
analysis using thinner (by a factor of 2) region sizes for the ``MINUS'' 
observation does not 
change the results presented here. We also avoid including the CCD borders maintaining a 
margin of $\sim$30$^{\prime\prime}$ from them. 

The results of the spectral fittings for the ``strip'' are shown in Table 2. Without any 
corrections for the overall gain difference between the 
pointings the regions with maximum velocity difference are {\bf R-2} with a best-fit redshift of 0.0084 and {\bf R+3} with 
z $\sim$ 0.0163 corresponding
to a velocity difference of $\sim$(2.4$\pm$1.0)$\times$10$^3$ km~s$^{-1}$, consistent with the 
maximum velocity differences found by {\sl ASCA}. These 
two regions have similar temperatures ($\sim$3.6 keV) and the iron abundance of the high velocity region is slightly 
lower (0.7 solar) than that of the low velocity region (solar).   

Although the velocity determination is highly driven by the FeK line at $\sim$ 6.7 keV, we can also determine 
velocities without the FeK line complex. Since the gain corrections are frequency dependent the best fit redshift 
determined from low energy spectral lines can be significantly different from that 
determined at high frequencies. However, for the cases analyzed here 
the differences for both regions were changed similarly so that the velocity {\it differences} were maintained.
For example, the regions {\bf R-2} and {\bf R+3} had their best-fit redshifts changed 
to (1.08$\pm$0.23)$\times$10$^{-2}$ and (1.83$\pm$0.30)$\times$10$^{-2}$, respectively, when the FeK region
was excluded from the spectral fittings. The resulting velocity difference (2.25$\pm$1.13)$\times$10$^3$ km~s$^{-1}$ is 
virtually identical to that derived with the inclusion of the FeK line complex.

\section{Regions of Maximum Velocity Gradient
                                        }
                                        
Despite the apparent overall gain shift between the observations and the difference in photon 
statistics we can estimate the velocity signals with highest statistical significance by dividing the 
difference of the best fit velocity from the average over the CCD by the error of the measured velocity.
We denote this error weighted deviation simply as deviation significance and we plot its
color contours in Figures 3. In 
Figures 3 the black and white represent negative and positive velocities, respectively, with respect to the 
CCD average velocity. 
The magnitude of the deviation significance shows how significant is the velocity structure. 
This allows us to probe the significance ``peaks'' for further analysis. 

From Figure 3a 
we can see that in the ``MINUS'' pointing there is a significant low redshift (blueshifted) zone that 
extends from the outer contours of the clusters' core ($>$1$^\prime$.5) towards the middle of the CCD 
($\sim$3$^\prime$--4$^\prime$ from the cluster core) significantly overlapping with the 
{\bf R-2} region and also with the low redshift zone P3 found with {\sl ASCA}
by DB01b. In Figure 3a we also see two relatively small redshifted regions on each side of 
the core of the cluster but at very small distances: one at the tip of the nuclear X-ray arm 
($<$1$^\prime$ from the core) and the other at $<$30$^{\prime\prime}$ South of the core. 
As mentioned previously, the proximity to the
cluster core reduces the reliability of the best fit velocities found with the methodology 
used here. Furthermore, 
the projected temperature in that region is very low 1.0--1.4 keV), 
and the FeK line complex is unseen. 
On the other hand, the redshift of the more extended blueshifted 
region near the CCD center is determined mostly 
from the FeK line given the higher temperatures (3--3.8 keV).
Since the spatial gain variations are in general frequency dependent, we should not 
compare directly redshifts determined from FeL with those from FeK lines, and conservatively 
we do not consider the velocity results from spectral fittings of regions that are two close ($<$1$^\prime$)
to the cluster's center. The ``PLUS'' pointing is in general more featureless but shows a relatively extended region of 
high velocities towards the SE of the cluster relatively near the detector's edge. The
significance of this  
positive velocity region is smaller than that of the blueshifted region in the ``MINUS'' pointing.

The upper limit for the velocity gradient in the Centaurus cluster can be estimated from a 
more detailed analysis of the 
regions surrounding the high significance peaks of Figures 3. We selected for that purpose two 
regions shown in Figures 1 as {\bf PEAK+} and {\bf PEAK-}. They are both rectangular regions 
2$^\prime$.3$\times$2$^\prime$.7 ({\bf PEAK-}) 
and 1$^\prime$.5$\times$2$^\prime$.5 ({\bf PEAK+}) located 2$^\prime$.7 ({\bf PEAK-}) and 3$^\prime$.7 ({\bf PEAK+}) 
away from the cluster's center, respectively, i.e., out of the 
core region. Individual spectral fittings of these two regions show mildly different temperatures 
and Fe abundance (Table 2). {\bf PEAK-} has a temperature of 3.43 keV and solar Fe abundance while {\bf PEAK+} 
is slightly hotter with a temperature of 3.80 keV and abundance 0.7 solar. The redshifts of these
two regions is significantly discrepant, as expected, and correspond to a velocity difference of 
(2.9$\pm$0.7) $\times$ 10$^3$ km~s$^{-1}$.
The 90\% and 99\% confidence contours for these two regions are shown in Figure 4a together with the line of equal 
redshifts. The F-test indicates that the redshifts are different at the 99.4\% confidence, where 
the $\chi^2$ of simultaneous spectral fittings of both regions with tied redshifts 
declines from 355.3 to 349.9 for 494
degrees of freedom when the redshifts of the two data groups are let free to vary. 

It should be noted that although these regions show the most significant velocity gradient observed 
in the two pointings {\bf PEAK+} does not overlap with the ``strip'' analyzed 
in the previous section that included regions {\bf R+1}, {\bf R+2} and {\bf R+3}. 
The region in that strip that is 
the closest to {\bf PEAK+} is {\bf R+1}. The large  
positive error bar of the best fit redshift measured for that region is due to a second 
minimum in $\chi^2$ space
near the same redshift value as {\bf PEAK+} and this suggests that this region may extend to  
and partially overlap with 
{\bf R+1}. Unfortunately, local temporal gain fluctuations 
are relatively high in {\bf R+1} (Figure 2c.) and, with the available observations, there are not enough photon 
statistics to disentangle multiple velocity components in the line
of sight spectroscopically. 

To determine the level of gain fluctuations necessary to blur the detected velocity difference found 
between {\bf PEAK-} and {\bf PEAK+} we used the Monte Carlo method. We performed 1000 simulations 
corresponding to {\bf PEAK-} and {\bf PEAK+} using the XSPEC tool FAKEIT. We input the same initial 
parameter values as those found for the real data except for the redshifts, which were 
initially set at some intermediary value (0.011). Background and responses correspondent
to the real observations were also included and Poisson errors were added. The results 
are shown in Figure 4b, where we plot the probability that the velocity difference 
between {\bf PEAK+} and {\bf PEAK-} is greater or equal than the value shown in the X-axis.
It can be seen that for the velocity gradient to be generated 
by chance (assumed here as $<$ 84\%) it would be necessary to have a $\gtrsim$5$\sigma$ gain fluctuation corresponding to an 
uncertainty of $\gtrsim$2500 km~s$^{-1}$ in the velocity difference (or $\gtrsim$0.6\% gain uncertainty 
per velocity measurement), which is significantly higher than that seen across the CCDs.

                                        \section{
Summary \& Discussion 
                                         }
In this work we report the analysis of velocity structures in the intracluster gas of 
the Centaurus cluster at spatial scales smaller than (but comparable to) that used in a previous {\sl ASCA} analysis.
The two off-center
pointings were taken in such a way as to minimize the temporal intrachip gain variations,
which is crucial when performing velocity tomography of the ICM with X-ray CCDs. Our analysis of the gain stability
suggests that the intrachip gain variability is higher than that measured before 2001 (Grant 2001),
limiting the precision of the measured velocity differences. 
We have confirmed the detection of significant a velocity gradient in the Centaurus cluster along a direction roughly 
perpendicular to the direction of Cen 45 subgroup. There is some uncertainty on the uniqueness of the direction of the 
velocity gradient and there are indications that the high velocity region may 
extended to the South.
The gradient is also seen in individual spectral lines belonging to FeL and FeK line groups
but with higher uncertainties.

Velocity gradients, both transitory or rotational,
can be caused by cluster mergers 
(e.g. Roettiger, Loken \& Burns 1997, Ricker 1998; Takizawa \& Mineshige 1998.) If the 
gradient is due to residual gas circulation around the cluster's 
center the corresponding circular velocity ($V_{circ}$) is (1.2$\pm$0.7)$\times$10$^3$~km~s$^{-1}$ as
derived from spectral fittings of the full spectra of regions {\bf R-2} and {\bf R+3}, where we include 
the local temporal gain variation in the errors.
This is consistent with 
sub(tran)sonic gas motion given by $\sqrt{\frac{5kT_{ICM}}{3\mu m_{p}}} \approx$ 
10$^3~(\frac{T_{keV}}{3.7 keV})^{(\frac{1}{2})}$~km~s$^{-1}$. This kind of configuration for the 
velocity gradient is further supported by the lack of other indications of
supersonic gas motions in the region analyzed, such as bow shocks.
Such circulation, either rotational or eddy-like (such as that predicted by Ricker \& Sarazin (2001)), 
is suggestive of past off-center mergers with a strong component 
near the line of sight towards Centaurus (Churazov et al. 1999; DB01b). The velocity difference
in the zones of maximum significance can be as high as (2.9$\pm$0.7) $\times$ 10$^3$ km~s$^{-1}$. However, 
these zones are discrepant in CCD coordinates and, therefore, spatial-temporal gain variations cannot be ruled out
as a factor for amplifying the velocity gradient. 
If the intrachip gain uncertainties (from Figure 2c) are incorporated in the velocity measurements in 
the regions of maximum significance, the errors in velocity difference would 
double ($\sim$ 2.5 $\times$ 10$^3$ km~s$^{-1}$).

From the measurements shown here we can estimate the relative 
importance of the velocity gradient as measured from {\bf R-2} and {\bf R+3}.
Assuming that the gas is uniformly rotating around the cluster's center at a distance R, the circulation time 
is\\  $\tau \sim 0.44 h_{50}^{-1}(\frac{R}{4^{\prime}.5})(\frac{V_{circ}}{1.2 \times 10^3~km~s^{-1}})^{-1}$Gyr.
In our particular case,
for a cylinder at a radial distance R and with height $\approx~\Delta$~R~$\approx$~R, the rotational 
energy can be roughly given by \\
$E_{rot}\sim 2\times 10^{61} h_{50}^{-3} (\frac{\eta}{2.5}) (\frac{\mu}{0.6})(\frac{n}{10^{-2}cm^{-3}})(\frac{V_{circ}}{1.2\times 10^{3}~km~s^{-1}})^2~(\frac{R}{4^{\prime}.5})^{3}~$ergs,
where n is the gas particle number density, $\mu$ is the mean molecular weight and $\eta$ is a 
geometric coefficient such that $\eta=\frac{I}{MR^2}$, where I is the moment of inertia and M the mass of the rotating body.  
Since the typical energy of cluster mergers is 10$^{63-64}$ ergs and considering that 
we are probing only a small region of the cluster our results imply that $>$0.1\%--1\% of the total merging kinetic 
energy can be still 
effectively converted into kinetic $\ga$~0.3--0.6 Gyr after the merging event. The ratio of bulk kinetic to thermal 
gas energies is given by 
$\beta_{gas} = 2 (\frac{\eta}{2.5})~(\frac{\mu}{0.6})(\frac{V_{circ}}{1.2\times~10^{3}~km~s^{-1}})^2~(\frac{kT}{3.7~keV})^{-1} \ga 1$,
implying that the X-ray derived mass within the region analyzed can be significantly underestimated if this high level of
departure from hydrostatic equilibrium is not taken into account.
%, for which $\beta_{gas}$=0
This work suggest that the intracluster gas can significantly depart from hydrostatic equilibrium 
even in cool clusters that do not show strong signs of merging such as Abell 3526.

%The presence of bulk motions in the cluster
%can bias the determination of cluster mass (and its derived parameters) and can induce core turbulence, helping to suppress
%radiative cooling near the core (e.g. Fujita et al. 2004).

The consistency between {\sl Chandra} \& {\sl ASCA} measurements of the velocity field 
illustrates the ability 
of current spectrometers to measure intracluster gas bulk motions with relatively 
short observations. Furthermore, with the demise of the 
calorimeter on-board {\sl Suzaku}, it is clear that 
the study of ICM velocity gradients depends heavily on 
multipointing observations made with different X-ray 
spectrometers, allowing us to reduce 
gain systematics and cross check the measurements. 

\acknowledgments We would like to thank E. Lloyd-Davies and J. Irwin  for the many helpful discussions 
and suggestions. We particularly thank Catherine Grant, George Chartas, Alexey Vikhlinin for providing helpful
information on calibration issues.
We acknowledge support from NASA Grant NAG 5-3247. This research made use of the HEASARC 
{\sl ASCA} database and NED.

%===============================================================================

%===============================================================================
%\begin{figure}
%\title{
Figure Captions
%}

\clearpage

\begin{figure}
\plotone{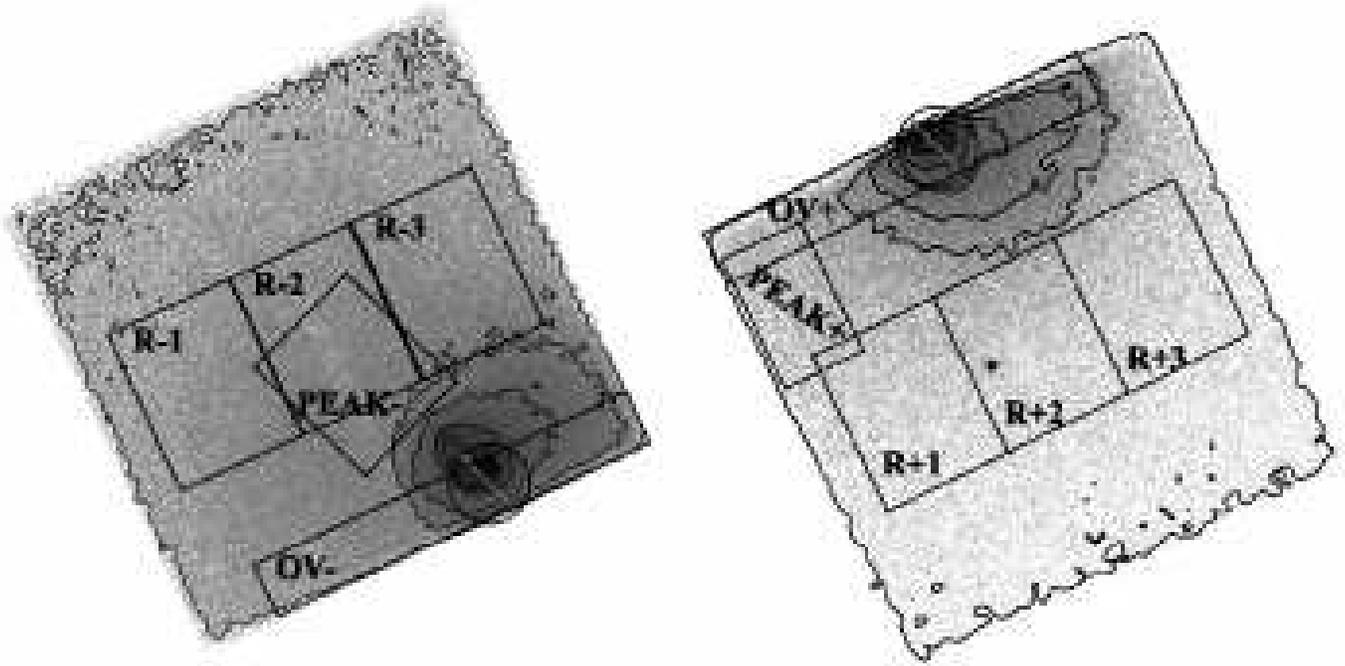}
\caption{
Extraction regions used for spectral fittings for the two pointings with ACIS-S3 analyzed in this work overlaid 
on the smoothed X-ray image. North is up for both pointings. (Left) ``MINUS'' pointing. The scale in RA is 
192.375$^{\circ}$--192.15$^{\circ}$ (left to right) and DEC -41.36$^{\circ}$-- -41.18$^{\circ}$ (top to bottom).
(Right) ``PLUS'' pointing. The scale in RA is 
192.30$^{\circ}$--192.05$^{\circ}$ (left to right) and DEC -41.28$^{\circ}$-- -41.45$^{\circ}$ (top to bottom).
The cluster's center is marked with a circle. 
X-ray contours are overlaid near the core and we also 
to show roughly the chip borders with a near zero surface brightness contour.}
\end{figure}  
                    
\clearpage

\begin{figure}
\vspace{-3in}
\epsscale{.80}
\plotone{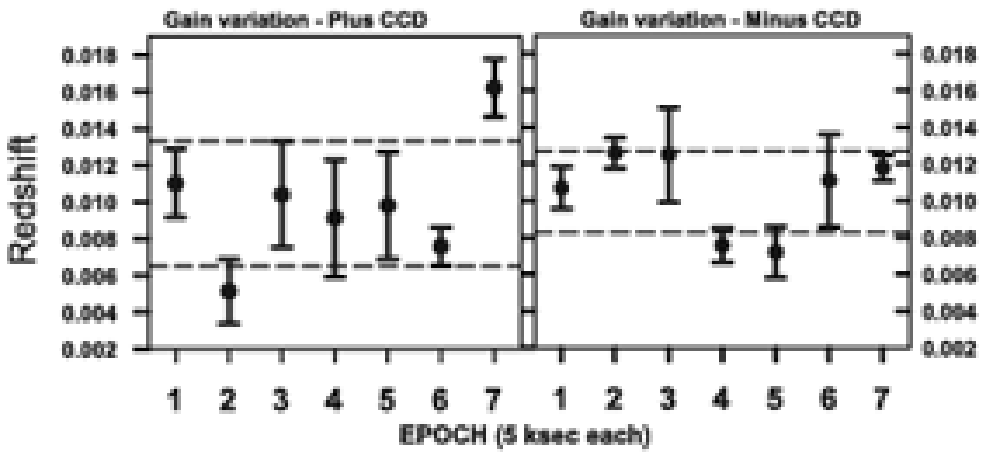}
\plottwo{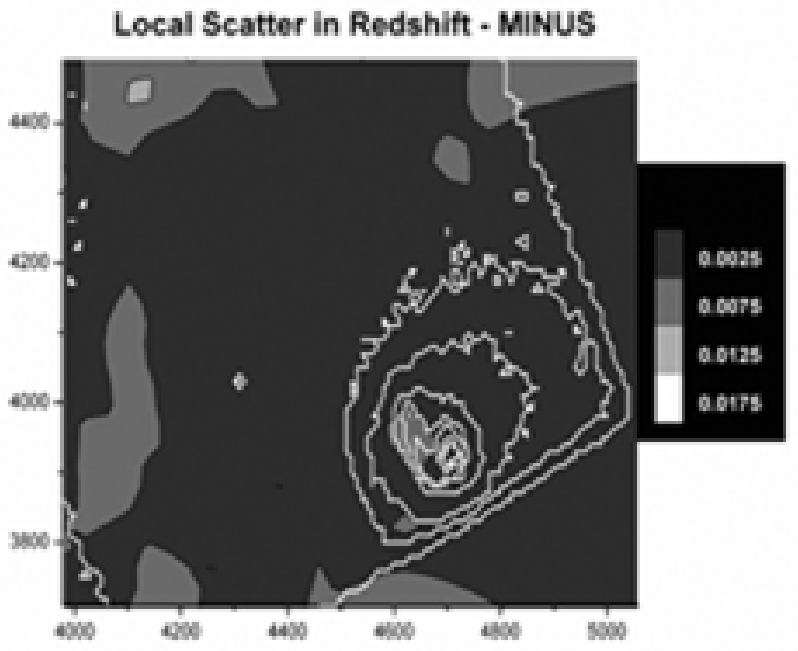}{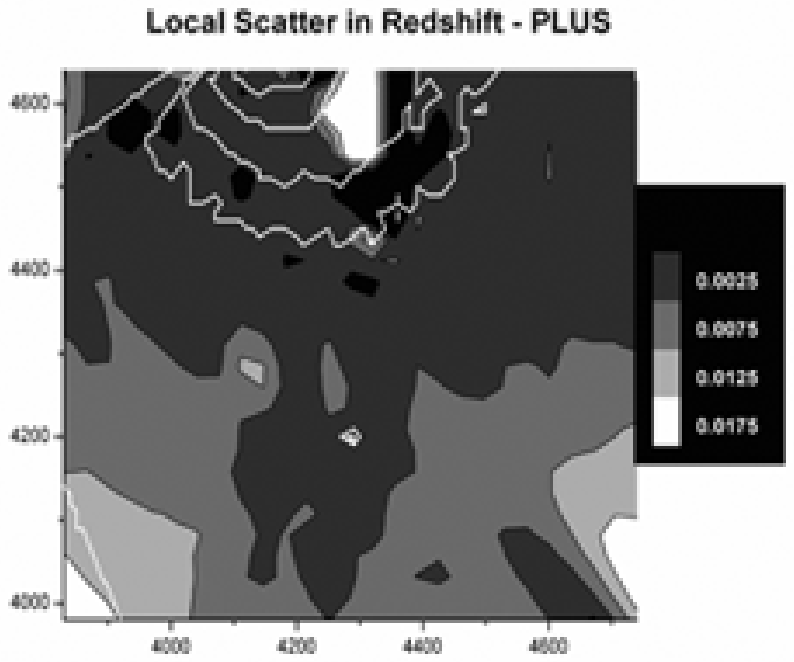}
\caption{
(a) Gain variation for the whole CCDs corresponding to the two observations 
(PLUS (Right) \& MINUS (Left)) in seven different epochs. The dashed black horizontal lines
show 1 standard deviation of the best-fit values. (b) and (c) Map of the standard 
deviation of the best-fit velocity found in each of the 7 epochs. The color scale step
represent the 1$\sigma$ fitting error for the measured redshift. The spatial scales of 
(b) and (b) are different because of the intrinsically lower brightness of the ``PLUS'' 
side and the fact that the velocities are found using an adaptive smoothing routine that
keeps the minimum number of counts fixed, therefore requiring larger region sizes for 
the ``PLUS'' side. The coloring outside the CCD borders are an artifact of the gridding 
routine and should be ignored.}
\end{figure}  

\clearpage

\begin{figure}
\plottwo{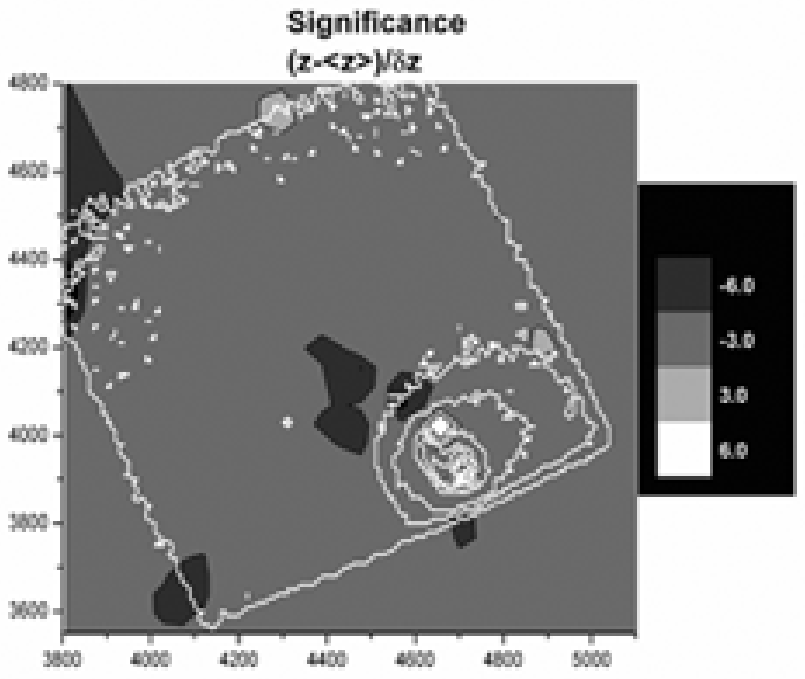}{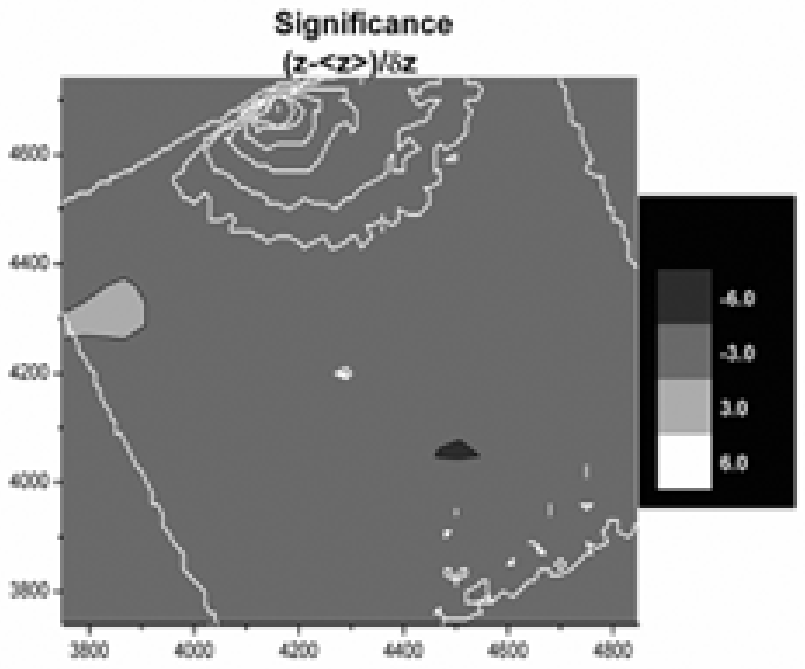}
\caption{
Results from an adaptive smoothing algorithm with a minimum of 5000 counts per extraction 
circular region and fitted with an absorbed VAPEC spectral model. The gridding method used 
is a correlation method that calculates a new value for each cell in the regular matrix from 
the values of the points in the adjoining cells that are included within the 
search radius. With the minimum count constraints the matrix 
size was 30 $\times$ 30 cells. We also overlay the X-ray contours shown in Figures 1 on top of the 
contour plot).}
\end{figure}  

\clearpage

\begin{figure}
\plottwo{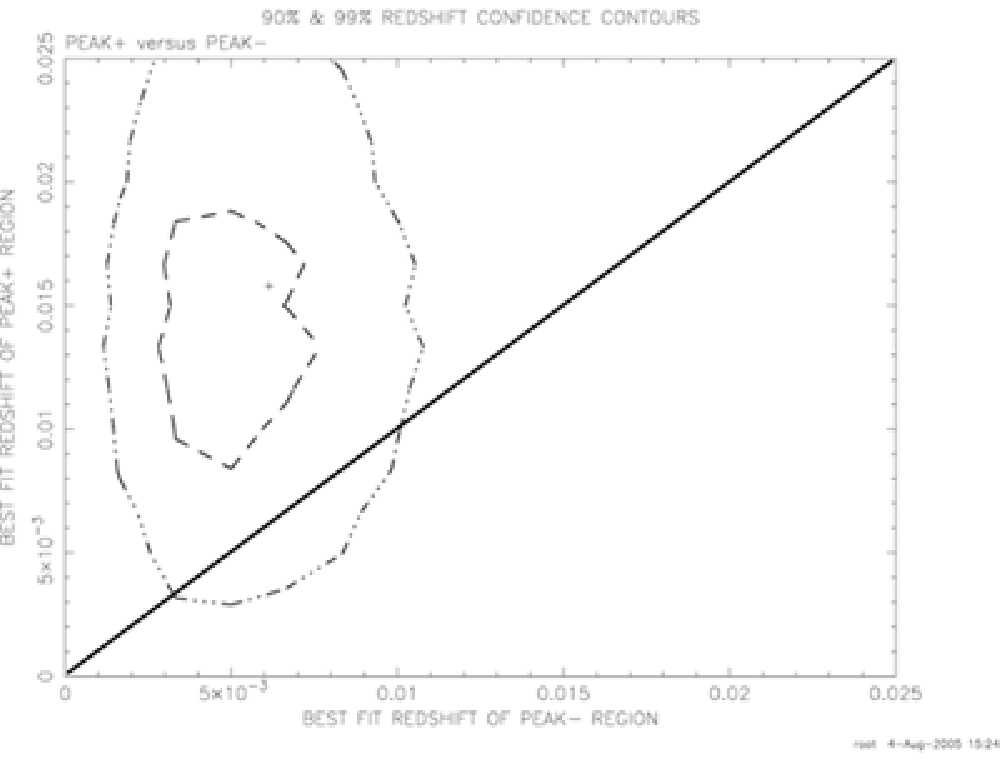}{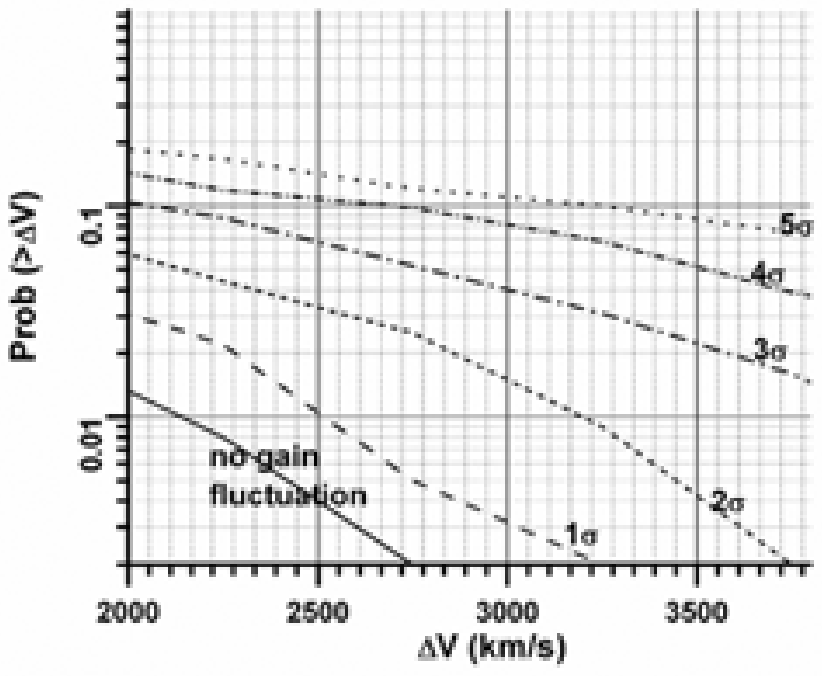}
\caption{
(a)Confidence contour plot for the redshifts measured for regions {\bf PEAK+} \& {\bf PEAK-} extraction region. 
The two contours 
correspond to 90\% (dashed) and 99\% (dotted) confidence levels. The 
contours are found from simultaneous 
spectral fittings of two data groups corresponding to the {\bf PEAK+} \& {\bf PEAK-} regions. We also 
indicate the line of equal redshifts.
(b) Probability of detecting a velocity difference greater than $\Delta$V for {\bf PEAK+} \& {\bf PEAK-} 
regions. Solid line is without gain fluctuations. 
The other plots assume a 1$\sigma$ (dashed), 2$\sigma$(short-dashed), 3$\sigma$(dash-dotted), 
4$\sigma$(short dash-dotted) and 5$\sigma$ (dotted)
gain fluctuation (1~$\sigma$ corresponds to 500km s$^{-1}$ for individual velocity differences). 
Results are obtained from spectral
fittings of 1000 simulated spectra for each region.}
\end{figure}  

\clearpage

%=== Table 1 ====================================================================

\begin{deluxetable}{lccccccc}
\tablewidth{0pt}
\tablecaption{Best fit Redshifts for the ``sky'' Overlapping Regions\tablenotemark{a,b}}
\tablehead{
\colhead{} &
\colhead{w/ core} &
\colhead{w/ core}  &
\colhead{w/ core } &
\colhead{w/out core } &
\colhead{w/out core} &
\colhead{w/out core} & \\
\colhead{} &
\colhead{0.5--8.5 keV} &
\colhead{1.0--8.5 keV} &
\colhead{0.5--5.0 keV} &
\colhead{0.5--8.5 keV} &
\colhead{1.0--8.5 keV} &
\colhead{0.5--5.0 keV} &
}
\startdata
z$_{OV+}$(10$^{-2}$)  & 1.285$\pm$0.015 & 0.91$\pm$0.016 & 1.29$\pm$0.025 & 0.895$\pm$0.02 & 0.91$\pm$0.04 & 1.175$\pm$0.075\\
z$_{OV-}$(10$^{-2}$)  & 2.256$\pm$0.07 & 1.525$\pm$0.015 & 2.20$\pm$0.02 & 1.30$\pm$0.02 & 1.287$\pm$0.03 & 1.54$\pm$0.015 \\
$\Delta$~V (10$^{3}$ km~s$^{-1}$) & 2.91$\pm$0.22 & 1.92$\pm$0.07 & 2.73$\pm$0.11 & 1.22$\pm$0.09 & 1.13$\pm$0.15 & 1.1$\pm$0.23 \\
$\chi~^2$/dof+ & 463/299 & 240/265 & 405/258 & 236/268 & 195/234 & 205/236\\
$\chi~^2$/dof- & 852/298	 & 238/264 & 817/258 & 220/263 & 166/229 & 187/233 \\
\enddata
\tablenotetext{a}{Errors are 1$\sigma$ confidence}
\tablenotetext{b}{Fits to a {\bf WABS VAPEC spectral} model}
%\tablenotetext{c}{Photospheric}
\end{deluxetable}

\clearpage

%=== Table 2 ====================================================================
\begin{deluxetable}{lccccc}
\small
\tablewidth{0pt}
\tablecaption{Spectral Line Fittings for Central CCD Regions of the ``PLUS'' \& ``MINUS'' Pointings\tablenotemark{a,b} }
\tablehead{
\colhead{Region} &
\colhead{Temperature} &
\colhead{Fe Abundance \tablenotemark{c}} &
\colhead{Redshift}  &
\colhead{$\chi^2$/dof} & \\
\colhead{} &
\colhead{(keV)} &
\colhead{(solar)} &
\colhead{(10$^{-2}$)} &
\colhead{} &
}
\startdata
{\bf R-1}  & 3.74$^{+0.11}_{-0.14}$ & 1.00$^{+0.13}_{-0.05}$ & 1.27$^{+0.05}_{-0.09}$ & 266/293\\ 
{\bf R-2}  & 3.58$^{+0.11}_{-0.11}$ & 0.98$^{+0.11}_{-0.07}$ & 0.84$^{+0.25}_{-0.14}$ & 255/317\\ 
{\bf R-3}  & 3.34$^{+0.10}_{-0.08}$ & 0.95$^{+0.10}_{-0.06}$ & 1.23$^{+0.08}_{-0.08}$ & 296/317\\ 
{\bf R+1}  & 3.79$^{+0.20}_{-0.20}$ & 0.55$^{+0.12}_{-0.08}$ & 0.92$^{+0.71}_{-0.19}$ & 203/253\\ 
{\bf R+2}  & 3.71$^{+0.16}_{-0.18}$ & 0.85$^{+0.13}_{-0.09}$ & 0.73$^{+0.15}_{-0.50}$ & 196/269\\ 
{\bf R+3}  & 3.61$^{+0.20}_{-0.22}$ & 0.67$^{+0.17}_{-0.09}$ & 1.63$^{+0.20}_{-0.34}$ & 193/259\\ 
{\bf {\bf PEAK-}}& 3.43$^{+0.09}_{-0.05}$ & 1.03$^{+0.05}_{-0.08}$ & 0.61$^{+0.05}_{-0.06}$ & 233/295\\ 
{\bf {\bf PEAK+}}& 3.80$^{+0.21}_{-0.22}$ & 0.71$^{+0.15}_{-0.14}$ & 1.58$^{+0.18}_{-0.30}$ & 118/198\\ 
\enddata
\tablenotetext{a}{Errors are 1$\sigma$ confidence}
\tablenotetext{b}{Full energy range (0.5 keV--8.5 keV)}
\tablenotetext{c}{Photospheric (Anders, \& Grevesse 1989)}
\end{deluxetable}

                               \end{document}